\documentclass{sig-alternate-2013}
\PassOptionsToPackage{hyphens}{url}
\usepackage{hyperref}

\begin{document}






%

\title{Persistent URIs Must Be Used To Be Persistent}
\numberofauthors{3} 
\author{
\alignauthor
Herbert Van de Sompel\\
       \affaddr{Los Alamos National Laboratory}\\
       \affaddr{Los Alamos, NM, USA}\\
       \affaddr{0000-0002-0715-6126}\\
       \email{herbertv@lanl.gov}
\alignauthor
Martin Klein\\
       \affaddr{University of California Los Angeles}\\
       \affaddr{Los Angeles, CA, USA}\\
       \affaddr{0000-0003-0130-2097}\\
       \email{martinklein@library.ucla.edu}
\alignauthor
Shawn M. Jones\\
       \affaddr{Los Alamos National Laboratory}\\
       \affaddr{Los Alamos, NM, USA}\\
       \affaddr{0000-0002-4372-870X}\\
       \email{smjones@lanl.gov}
}
\maketitle
\begin{abstract}
We quantify the extent to which references to papers in scholarly literature use persistent HTTP URIs  
that leverage the Digital Object Identifier infrastructure. We find a significant number of references 
that do not, speculate why authors would use brittle URIs when persistent ones are available, 
and propose an approach to alleviate the problem.
\end{abstract}
%
%
%
%
%
%
%
%
%
\keywords{Digital Object Identifier, Scholarly Communication, URI References}
\section{Introduction and Motivation}
Motivated by a desire to achieve persistence when linking to web resources, 
various solutions have been introduced aimed at decoupling the identification 
and location of a web resource by means of HTTP URIs; a comprehensive overview is available 
from \cite{duerr:identification}.
Solutions such as PURL, W3ID, identifiers.org, and DOI make use of a resolver 
infrastructure to achieve this goal. They use an \textbf{identifying HTTP URI} to persistently 
identify a web resource, and a \textbf{locating HTTP URI} for the resource's current web location.
The custodian of a web resource 
maintains the correspondence between the identifying URI and the locating URI 
in the resolver's look-up table as the resource's location changes over time. 
When a client accesses the identifying URI, it is redirected to the locating URI.

This solution is attractive, especially when it is to be expected that the domain where a resource is located 
may change over time. This is, for example, the case with academic journals that move hands between publishers 
as acquisitions and mergers take place. The solution comes at a price because it requires operating a resolver infrastructure 
and maintaining the look-up table that powers it. But, as long as the identifying URI is used to link to a resource, 
the solution achieves its goal of link persistence. 

In 2014, we conducted a large-scale study about reference rot\footnote{\url{http://mementoweb.org/missing-link/}} 
in web-based scholarly communication \cite{klein:one_in_five}. The focus of our study was on links to so-called 
\textbf{web-at-large} resources found in scholarly articles, that is, 
links to resources that are themselves not scholarly papers. Our intuition suggested that, 
in order to filter out links to scholarly articles, we only would have to remove links 
targeted at the DOI resolver, i.e. with baseURL \texttt{dx.doi.org}. After all, academic publishers 
started assigning DOIs to papers about two decades ago, and the practice has been the norm across disciplines for many years. 
However, when eyeballing the result of this DOI-based filtering, we were stunned to be left with many links with baseURLs that 
were clearly associated with academic publishers, for example \texttt{biomedcentral.com}, \texttt{sciencedirect.com}, and
\texttt{link.springer.com}. To put it differently, we found a significant number of references to papers linked by their locating URI 
instead of their identifying URI. For these links, the persistence intended by the DOI persistent identifier infrastructure 
was not achieved. In this poster, we determine how widespread this problem is, we speculate on the origin of the problem, 
and propose a possible way to address it.
\section{Methodology}
For the aforementioned reference rot study, we collected more than $1.8$ million papers published between $1997$ and $2012$, 
in three scholarly corpora: the entire arXiv.org preprint collection, a sizable sample of papers in Elsevier journals, 
and all papers submitted to PubMed Central. 
%
We extracted almost $4$ million URI references using advanced regular expression 
techniques \cite{zhou:no_more_404} from all sections 
of those papers including footnotes, tables, and references.
We dismissed $1.7$ million URI references because they were to license statements 
such as \texttt{creativecommons.org/licenses/} or to miscellaneous resources such as \texttt{127.0.0.1} 
and \texttt{www.example.org}.
The data is available via \cite{klein:one_in_five}.

For this poster, we take the remaining, approximate $2.2$ million, 
URI references as a starting point, 
and use them to determine how commonly scholarly papers are referenced by means of their locating URIs instead of 
their persistent, DOI-based, identifying URIs. We observe that Elsevier papers hardly contain any 
DOI references, an artefact of the formatting of papers obtained via the 
CrossRef Text and Data Mining API\footnote{\url{http://tdmsupport.crossref.org/}}; DOI references are added on-the-fly by 
Elsevier's dissemination platform. This observation leads us to exclude the Elsevier corpus from this study. 
We proceed to categorize the remaining $1.6$ million URI references from arXiv and PubMed Central as follows:
\begin{itemize}
\item \textbf{DOI} : Scholarly paper referenced by means of an identifying DOI-based URI - Selecting these references is trivial 
as they all have \texttt{dx.doi.org} as their baseURL. 
\item \textbf{shouldBeDOI}: Scholarly paper referenced by means of a locating URI - In order to select these references, we use 
a list of hash values of publisher baseURLs provided by CrossRef\footnote{\url{http://labs.crossref.org/reverse-domain-lookup/}}.
If the hash of a baseURL of an 
extracted reference matches a hash in CrossRef's list, a reference is added to this category.
\item \textbf{web-at-large}: Web resource referenced -- references that 
do not fall into the DOI nor the shouldBeDOI category.
\end{itemize}
\begin{figure}[t!]
\center
\includegraphics[scale=0.25]{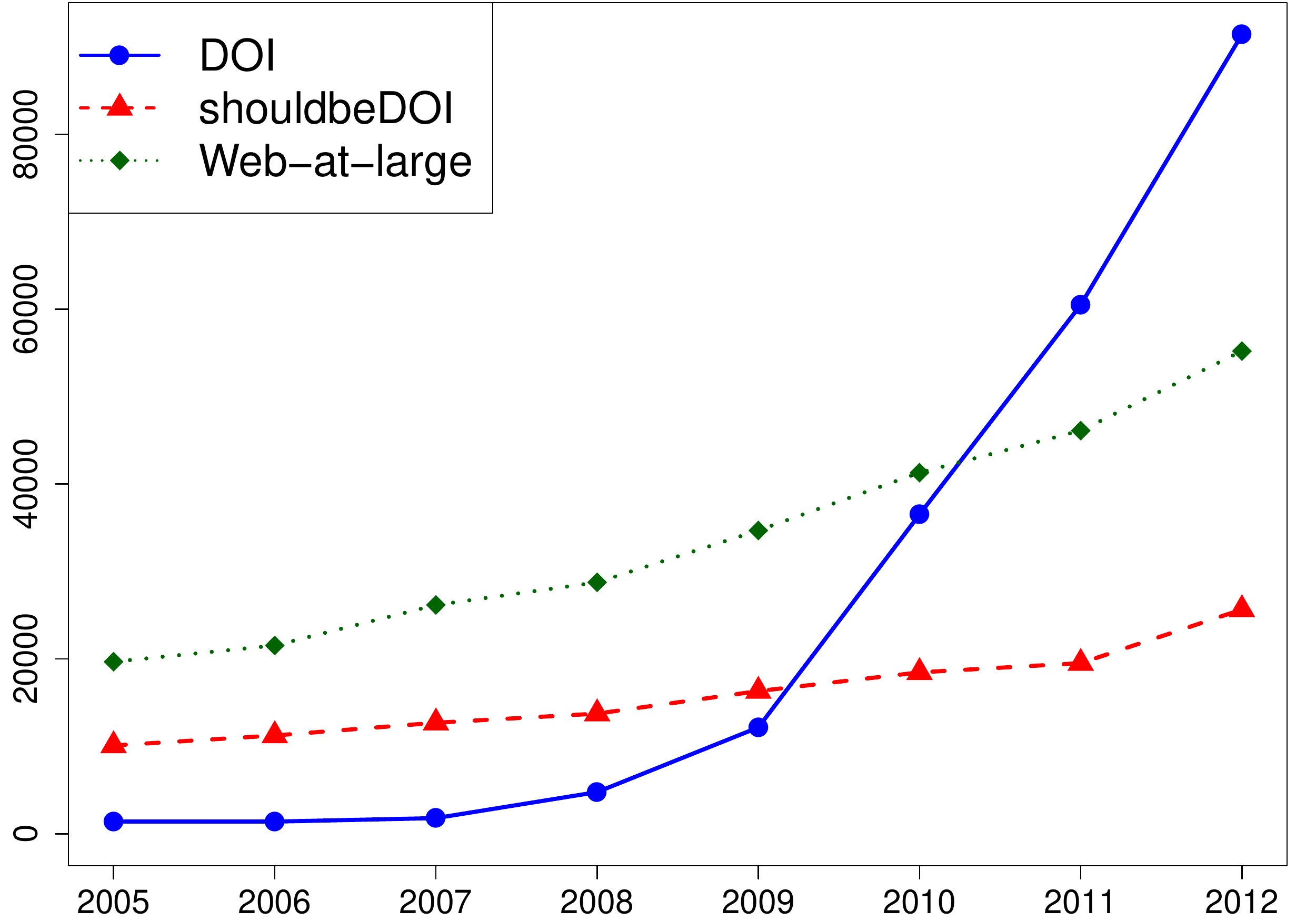}
\caption{arXiv}
\label{fig:arxiv_refs}
\end{figure}
\section{Results}
\begin{figure}[t!]
\center
\includegraphics[scale=0.25]{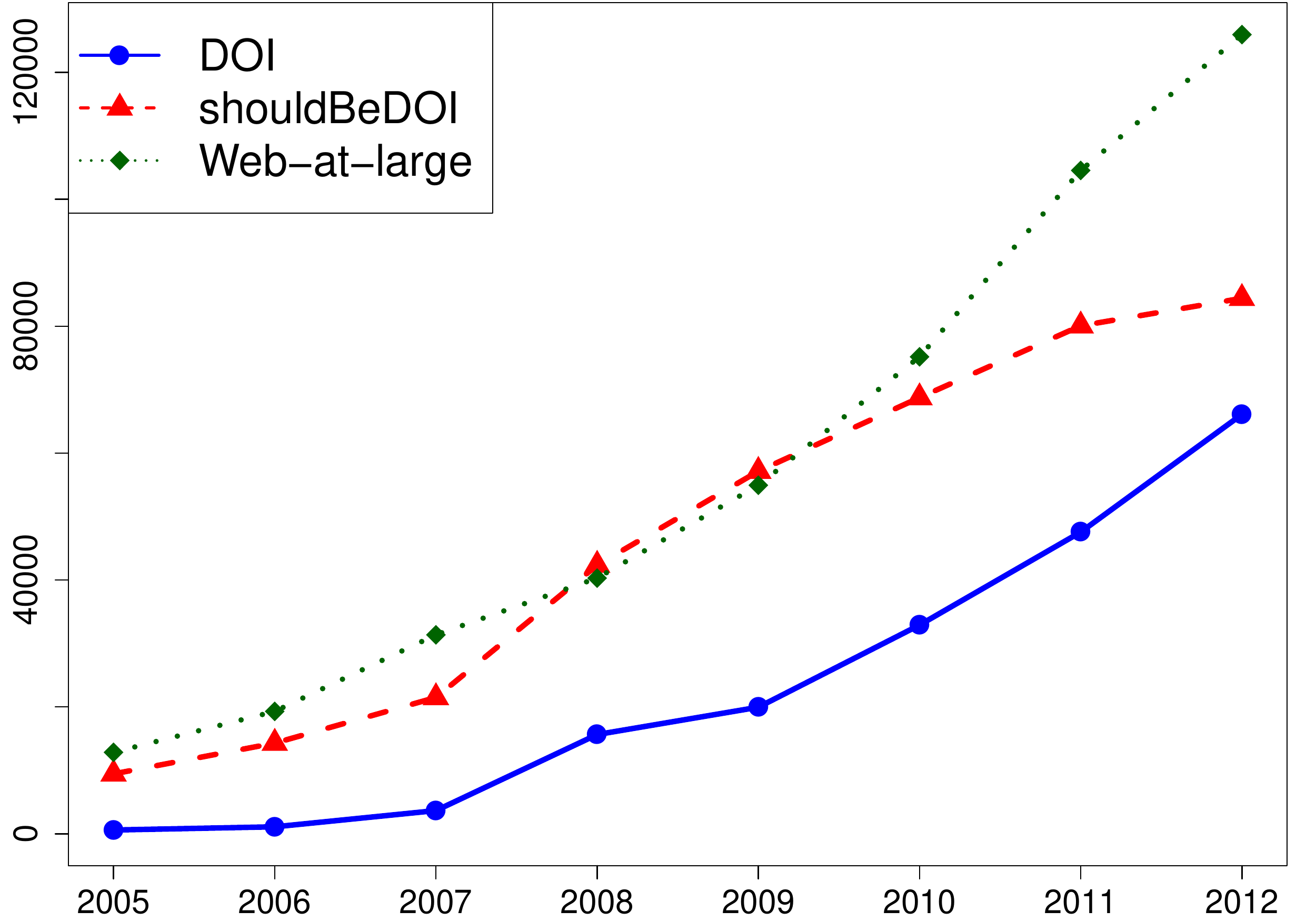}
\caption{PMC}
\label{fig:pmc_refs}
\end{figure}

Following the aforementioned methodology, we end up with $397,412$ DOI, $505,647$ shouldBeDOI, and $737,847$
web-at-large references. We depict the distribution of these references in function of the 
publication date of the referencing paper in Figure \ref{fig:arxiv_refs} for arXiv 
and Figure \ref{fig:pmc_refs} for PubMed Central. DOI references are blue, shouldBeDOI red, and web-at-large green. 
The publication date on the x-axis ranges from $2005$ to $2012$. 

In both figures, each category grows over time, consistent with the continuous growth of 
paper publications per year and the increased use of HTTP URI references \cite{klein:one_in_five}. 
In the arXiv corpus, shouldBeDOI grows slowly and steadily whereas DOI makes a sudden jump around $2009$, 
possibly related to the provision of DOIs in downloadable references for Physics, Mathematics, and Statistics, 
which constitute the majority of this corpus. 
From $2009$ onwards, DOI outnumbers shouldBeDOI but a significant number of shouldBeDOI references remain, for example, 
about $20,000$ in $2012$. The pattern in PubMed Central is quite different. Growth for all categories 
kicks off in $2008$, consistent with NIH submission 
mandates\footnote{\url{http://sciencecommons.org/weblog/archives/2009/03/17/nih-mandate-made-permanent/}}. 
But, for all publication years, shouldBeDOI outnumbers DOI, and web-at-large tops both. An astonishing number of 
papers are referenced by means of their location URI instead of their DOI-based identifying URI, for example, 
about $80,000$ in $2012$.

These results must be interpreted subject to two caveats. First, the list of hash values of publisher baseURLs 
is an approximation as it represents the current state of location URIs in the CrossRef resolver. As a result, some 
shouldbeDOI references may have been categorized as web-at-large because a location URI used years ago 
is no longer in use today. Second, lacking public information as to when publishers started assigning DOIs to papers, 
some references may have wrongly been categorized as shouldbeDOI because a publisher did not yet use DOIs prior to 
a certain publication date. However, given the widespread adoption of DOIs by scholarly publishers in recent years, 
we feel confident that our analysis is reliable for the depicted timeframe.
\section{Discussion and Conclusions}
We quantified the widespread use of location URIs instead of DOI-based identifying URIs for referencing papers in 
scholarly literature. We do not know why authors choose brittle links over persistent ones. But, 
being authors ourselves, we speculate that it has to do with the use of the location URI when bookmarking and 
creating entries in citation management tools. After all, the paper and associated information is available at 
the location URI, not the identifying URI. In order to alleviate this problem, we propose the use of a 
Link\footnote{\url{http://www.ietf.org/rfc/rfc5988.txt}} header field in the 
HTTP response of the location URI that conveys a link pointing with an appropriate relation type 
- \textbf{canonical} comes to mind - to the identifying URI. 
Bookmarks and citation managers could use this Link information to record the identifying URI and make links that were 
intended to be persistent actually persistent.
%

\begin{thebibliography}{1}

\bibitem{duerr:identification}
R.~Duerr and R.~{Downs et al.}
\newblock {On the Utility of Identification Schemes for Digital Earth Science
  Data: An Assessment and Recommendations}.
\newblock {\em Earth Science Informatics}, 4(3):139--160, 2011.

\bibitem{klein:one_in_five}
M.~Klein and H.~{Van de Sompel et al.}
\newblock {Scholarly Context Not Found: One in Five Articles Suffers from
  Reference Rot}.
\newblock {\em PLoS ONE}, 9(12), 2014.

\bibitem{zhou:no_more_404}
K.~Zhou and C.~{Grover et al.}
\newblock {No More 404s: Predicting Referenced Link Rot in Scholarly Articles
  for Pro-Active Archiving}.
\newblock In {\em Proceedings of JCDL '15}, pages 233--236, 2015.

\end{thebibliography}
%
%

%
%
\end{document}